\documentclass[longauth]{aa} 

\usepackage{graphicx}
\usepackage{txfonts}
\begin{document}

\title{First detection of the Crab Nebula at TeV energies with a Cherenkov telescope in a dual-mirror Schwarzschild-Couder configuration: the ASTRI-Horn telescope}
\author{
  S.~Lombardi\inst{1,2,\star} \and
  O.~Catalano\inst{3,\star} \and
  S.~Scuderi\inst{4,\star} \and
  L.~A.~Antonelli\inst{1,2} \and
  G.~Pareschi\inst{5} \and
  E.~Antolini\inst{6} \and
  L.~Arrabito\inst{7} \and
  G.~Bellassai\inst{8} \and
  K.~Bernl\"{o}hr\inst{9} \and
  C.~Bigongiari\inst{1} \and
  B.~Biondo\inst{3} \and
  G.~Bonanno\inst{8} \and
  G.~Bonnoli\inst{5} \and
  G.~M.~B\"{o}ttcher\inst{10} \and
  J.~Bregeon\inst{11} \and
  P.~Bruno\inst{8} \and
  R.~Canestrari\inst{3} \and
  M.~Capalbi\inst{3} \and
  P.~Caraveo\inst{4} \and
  P.~Conconi\inst{5} \and
  V.~Conforti\inst{12} \and
  G.~Contino\inst{3} \and
  G.~Cusumano\inst{3} \and
  E.~M.~de~Gouveia Dal Pino\inst{13} \and
  A.~Distefano\inst{4} \and
  G.~Farisato\inst{14} \and
  C.~Fermino\inst{13} \and
  M.~Fiorini\inst{4} \and
  A.~Frigo\inst{14} \and
  S.~Gallozzi\inst{1} \and
  C.~Gargano\inst{3} \and
  S.~Garozzo\inst{8} \and
  F.~Gianotti\inst{12} \and
  S.~Giarrusso\inst{3} \and
  R.~Gimenes\inst{13} \and
  E.~Giro\inst{14} \and
  A.~Grillo\inst{8} \and
  D.~Impiombato\inst{3} \and
  S.~Incorvaia\inst{4} \and
  N.~La~Palombara\inst{4} \and
  V.~La~Parola\inst{3} \and
  G.~La~Rosa\inst{3} \and
  G.~Leto\inst{8} \and
  F.~Lucarelli\inst{1,2} \and
  M.~C.~Maccarone\inst{3} \and
  D.~Marano\inst{8} \and
  E.~Martinetti\inst{8} \and
  A.~Miccich\`e\inst{8} \and
  R.~Millul\inst{5} \and
  T.~Mineo\inst{3} \and
  G.~Nicotra\inst{15} \and
  G.~Occhipinti\inst{8} \and
  I.~Pagano\inst{8} \and
  M.~Perri\inst{1,2} \and
  G.~Romeo\inst{8} \and
  F.~Russo\inst{3} \and
  F.~Russo\inst{12} \and
  B.~Sacco\inst{3} \and
  P.~Sangiorgi\inst{3} \and
  F.~G.~Saturni\inst{1} \and
  A.~Segreto\inst{3} \and
  G.~Sironi\inst{5} \and
  G.~Sottile\inst{3} \and
  A.~Stamerra\inst{1} \and
  L.~Stringhetti\inst{4} \and
  G.~Tagliaferri\inst{5} \and
  M.~Tavani\inst{16} \and
  V.~Testa\inst{1} \and
  M.~C.~Timpanaro\inst{8} \and
  G.~Toso\inst{4} \and
  G.~Tosti\inst{17} \and
  M.~Trifoglio\inst{12} \and
  G.~Umana\inst{8} \and
  S.~Vercellone\inst{5} \and
  R.~Zanmar~Sanchez\inst{8} \and
  C.~Arcaro\inst{14} \and
  A.~Bulgarelli\inst{12} \and
  M.~Cardillo\inst{16} \and
  E.~Cascone\inst{18} \and
  A.~Costa\inst{8} \and
  A.~D'A\`i\inst{3} \and
  F.~D’Ammando\inst{12} \and
  M.~Del~Santo\inst{3} \and
  V.~Fioretti\inst{12} \and
  A.~Lamastra\inst{1} \and
  S.~Mereghetti\inst{4} \and
  F.~Pintore\inst{4} \and
  G.~Rodeghiero\inst{14} \and
  P.~Romano\inst{5} \and
  J.~Schwarz\inst{5} \and
  E.~Sciacca\inst{8} \and
  F.~R.~Vitello\inst{8} \and
  A.~Wolter\inst{5}
}
\institute {
  INAF Osservatorio Astronomico di Roma, 00078 Monte Porzio Catone (Roma), Italy \and 
  ASI Space Science Data Center, 00133 Roma, Italy \and 
  INAF IASF Palermo, 90146 Palermo, Italy \and 
  INAF IASF Milano, 20133 Milano, Italy \and 
  INAF Osservatorio Astronomico di Brera, 23807 Merate (LC), Italy \and 
  CTA Observatory, 40129 Bologna, Italy \and 
  LUPM IN2P3, Universit\'e de Montpellier, 34095 Montpellier, France \and 
  INAF Osservatorio Astronomico di Catania, 95123 Catania, Italy \and 
  MPI f\"{u}r Kernphysik, 69117 Heidelberg, Germany \and 
  CSR, North-West University, 2520 Potchefstroom, South Africa \and 
  LPSC IN2P3, Universit\'e de Grenoble-Alpes, 38400 Saint-Martin-d'H\`eres, France \and 
  INAF OAS Bologna, 40129 Bologna, Italy \and 
  IAG USP, Universidade de S{\~a}o Paulo, 5508-090 S{\~a}o Paulo, SP, Brazil \and 
  INAF Osservatorio Astronomico di Padova, 35122 Padova, Italy \and 
  INAF IRA Noto station, 96017 Noto (SR), Italy \and 
  INAF IAPS Roma, 00133 Roma, Italy \and 
  Dipartimento di Fisica e Geologia, Universit\`a di Perugia, 06123 Perugia, Italy \and 
  INAF Osservatorio Astronomico di Capodimonte, 80131 Napoli, Italy 
}

\date{Received 26 September 2019~/~Accepted 20 December 2019}

\abstract
{
We report on the first detection of very high-energy (VHE) gamma-ray emission from the Crab Nebula by a Cherenkov telescope in dual-mirror Schwarzschild-Couder (SC) configuration. This result has been achieved by means of the 4-m ASTRI-Horn telescope, operated on Mt. Etna, Italy, and developed in the context of the Cherenkov Telescope Array Observatory preparatory phase.
The dual-mirror SC design is aplanatic and characterized by a small plate scale, which allows us to implement large cameras with a large field of view, with small-size pixel sensors and a high level of compactness. The curved focal plane of the ASTRI camera is covered by silicon photo-multipliers (SiPMs), managed by an unconventional front-end electronic system that is based on a customized peak-sensing detector mode. The system includes internal and external calibration systems, hardware and software for control and acquisition, and the complete data archiving and processing chain.
These observations of the Crab Nebula were carried out in December 2018 during the telescope verification phase for a total observation time (after data selection) of 24.4~h, equally divided between on- and off-axis source exposure. The camera system was still under commission and its functionality was not yet completely exploited. Furthermore, due to recent eruptions of the Etna Volcano, the mirror reflection efficiency was reduced. Nevertheless, the observations led to the detection of the source with a statistical significance of 5.4~$\sigma$ above an energy threshold of $\sim$3~TeV.
This result provides an important step toward the use of dual-mirror systems in Cherenkov gamma-ray astronomy. A pathfinder mini-array based on nine ASTRI-like telescopes with a large field-of-view is in the course of implementation.
}


\titlerunning{First detection of Crab Nebula at VHE with a SC Cherenkov telescope}

\authorrunning{Lombardi~et~al.}

\maketitle

\section{Introduction}
\label{sec:section1}
%
\begingroup
\let\thefootnote\relax\footnotetext
    {$\star$ Corresponding authors:\\
      S.~Lombardi ({\tt saverio.lombardi@inaf.it}), O.~Catalano ({\tt osvaldo.catalano@inaf.it}), and S.~Scuderi ({\tt salvatore.scuderi@inaf.it}).
    }
\endgroup
Gamma-ray observation using the imaging atmospheric Cherenkov technique (IACT) (see e.g., \citealt{denaurois15} for a review) has experienced rapid growth in the past years. Exactly 30 years ago (1989), the very first detection of the Crab Nebula at TeV energies was achieved by the Whipple telescope~\citep{weekes89}. Since then, several experiments using IACT have been carried out, truly opening up the TeV astronomy era~\citep{hinton09} and leading to the detection of about 200 gamma-ray sources~\citep{tegevcat}. In particular, the latest generation of IACT observing facilities (i.e., H.E.S.S., MAGIC, and VERITAS) have allowed for unprecedented insights into the non-thermal very high-energy (VHE) universe~\citep{aharonian13,acharya19}. For the optical design, past and current generations of IACT experiments have adopted either the Davies-Cotton~\citep{davies57} or parabolic configurations~\citep{bastieri05}, using a single-mirror and a camera located at the focus of the telescope. Single-mirror systems in IACT astronomy are limited in terms of their field of view (since these systems are not aplanatic and greatly suffer from off-axis aberrations) and make use of cameras that are in general very large and bulky due to intrinsic large plate-scale. Innovative optical designs such as the dual-mirror Schwarzschild-Couder (SC) configuration, initially proposed for ground-based gamma-ray astronomy by~\cite{vassiliev07}, can represent a step forward for Cherenkov telescopes. This is mainly because they allow us to implement a larger field of view (up to 10$^{\circ}$ in diameter) in a much compact instrument since the plate-scale is much smaller. As a consequence, the adoption of small pixel size (6-7~mm linear dimension) silicon photo-multipliers (SiPMs) is possible as an alternative to traditional photo-multiplier tubes (PMTs).

\begin{figure}
\centering
\includegraphics[width=0.3\textwidth]{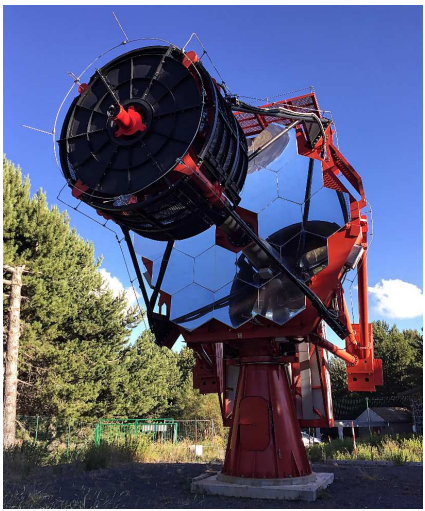}
\caption{ASTRI-Horn dual-mirror Cherenkov telescope installed on Mt. Etna (Italy) at the INAF M.C. Fracastoro observing station.}
\label{fig:figure1}
\end{figure}

Within the framework of the Cherenkov Telescope Array (CTA) Observatory Project, the Italian National Institute for Astrophysics (INAF) is leading the ASTRI ({\it Astrofisica con Specchi a Tecnologia Replicante Italiana}) Project~\citep{pareschi16,scuderi18}. The primary goal of the project has been the design, development, and deployment of an end-to-end 4-m diameter prototype of the CTA small-size telescopes (SSTs) with a dual-mirror configuration. The CTA SSTs sub-array will be composed by 70 telescopes and installed at the CTA Observatory southern site, with the aim to investigate the gamma-ray sky between a few~TeV and up to 100~TeV and beyond. The ASTRI prototype (see Fig.~\ref{fig:figure1}), named the ASTRI-Horn telescope~--~in honor of the Italian-Jewish astronomer Guido Horn D'Arturo, who pioneered the use of segmented primary mirrors in astronomy~\citep{horn36,jacchia78}~--~, is located on Mt. Etna (Italy) at the INAF M.C. Fracastoro observing station (37.7$^\circ$N, 15.0$^\circ$E, 1740~m a.s.l.)~\citep{maccarone13} and has undergone a verification phase over the past two years.
The ASTRI-Horn telescope is based upon a wide-field dual-mirror optical design~\citep{canestrari13,sironi17}, inspired by the SC aplanatic configuration~\citep{schwarzschild1905,couder1926,vassiliev07}. Also see~\citealt{lynden-bell02} for a review. It was also inspired by an innovative SiPM curved-focal plane camera~\citep{catalano18} managed by very fast read-out electronics~\citep{sottile16}. ASTRI includes, in addition to all other sub-systems~\citep{maccarone17}, full data acquisition~\citep{conforti16}, archiving~\citep{carosi16}, and processing~\citep{lombardi16} chain, from raw data up to final scientific products. For this purpose, dedicated software for the reduction and scientific analysis of the ASTRI data has been developed. The software has been extensively checked on a Monte Carlo (MC) basis and is currently exploited for the processing of data taken by the ASTRI-Horn telescope~\citep{lombardi17,lombardi18}. In order to test the performance of the ASTRI-like telescope layout in an array configuration, a mini-array composed of nine ASTRI telescopes is being developed and operated by INAF in the context of the preparatory effort for participating in CTA. The ultimate goal of the ASTRI Project is to contribute to the installation and operation of a considerable number of telescopes out of the 70 foreseen for CTA SSTs.

It should be noted that in CTA, other groups are also involved in the development of telescope prototypes based on dual-mirror SC configuration and compact cameras based on SiPMs. Among them, it is worth mentioning the Gamma-ray Cherenkov Telescope (GCT, another SST prototype,~\citealt{leblanc18}) and the prototype Schwarzschild-Couder telescope (pSCT, a $\sim$10~m diameter prototype for the medium-size telescopes array of CTA,~\citealt{rousselle15}). Also, other aplanatic configurations for wide-field Cherenkov telescopes, different from the SC design, have been proposed. For example,~\cite{mirzoyan09} and later,~\cite{cortina16} proposed a solution based on the modified Schmidt configurations with the use of a Fresnel type correction lens and a curved focal plane in addition to an aspherical reflecting mirror, while~\cite{maccarone08}  explored the use of Fresnel lenses in the context of the Gamma Air Watch (GAW) Project.

The verification and scientific validation of any ground-based gamma-ray instrument are typically performed by observing well-known and bright gamma-ray sources. In the case of the ASTRI-Horn telescope, the Crab Nebula, the remnants of a supernova explosion that occurred in AD 1054 at a distance of $\sim$2~kpc, has been considered. This source is in fact one of the best studied celestial objects emitting in almost all wavelength bands of the electromagnetic spectrum, from radio to gamma rays. It is the strongest source of steady VHE gamma-ray emission and is considered as the standard ``calibration candle'' for many ground-based gamma-ray instruments.

In this work, we present an overall description of the ASTRI-Horn telescope and its main sub-systems. In the subsequent sections, we present the data analysis and results of the observations of the Crab Nebula carried out in December 2018.

\section{The ASTRI-Horn telescope}
\label{sec:section2}
The ASTRI-Horn telescope has been developed in the context of prototyping the SSTs for the CTA Observatory Project. Its main technological innovation consists of the optical system, which implements the dual-mirror SC configuration. The primary mirror, 4.3~m in diameter, is composed of an array of 18 hexagonal tiles, while the secondary mirror, 1.8~m in diameter, consists of a monolithic hemispherical thick glass shell thermally bent to 2.2~m radius of curvature~\citep{canestrari13}. The telescope-equivalent focal length is 2.15~m (f/0.5) and the system covers a full field of view (FoV) of more than 10$^{\circ}$~\citep{rodeghiero16}. The SC optical design for Cherenkov telescopes has been fully validated for the first time by means of the ASTRI-Horn telescope~\citep{giro17}. Images of the Polaris Star at different off-axis angles (up to 4.5$^{\circ}$) have been taken with an optical CCD camera able to scan the whole FoV, demonstrating that the point spread function (PSF) has a constant width of a few arcmin over the whole FoV.

Since the SC optical design typically results in a small plate scale (37.5~mm/$^{\circ}$ in the case of ASTRI), the camera at the focal plane is characterized by compact dimensions over a large FoV. For this, SiPM sensors represent an optimal solution, as they have a typical lateral size of a few millimeters, very fast response, high photon detection efficiency, and excellent single photo-electron resolution~\citep{bonanno16,romeo18}. This class of detectors has been already demonstrated to be suitable in the fields of high-energy astrophysics and IACT applications, for example, by the single-mirror FACT telescope~\citep{anderhub13}. In the ASTRI camera~\citep{catalano18}, the SiPM sensors are organized in 37 photon detection modules (PDMs), each one with 8$\times$8 pixels. Each pixel has a linear dimension of 7~mm and an angular size of 0.19$^{\circ}$. Currently, the camera of the ASTRI-Horn telescope includes 21 out of 37 PDMs, for a total effective FoV of 7.6$^{\circ}$. This configuration matches the angular resolution of the optical system, providing the PSF (defined as the 80\% of the light collected from a point-like source) is contained in less than one camera pixel. The camera electronics is based on a custom peak detector to acquire the SiPM pulses~\citep{sottile16}. The Cherenkov imaging telescope-integrated read-out chip (CITIROC) has a signal shaper and peak detector customized for ASTRI. It represents an innovative technical solution and provides high efficiency pixel by pixel trigger capability, very fast camera pixel read~out~\citep{impiombato15}, and a dynamic range up to 1500~photo-electrons (pe). The trigger of the ASTRI camera is topological. It is activated when a given number of contiguous pixels within a PDM measures a signal above a given threshold. Both the number of contiguous pixels required for the trigger and the signal threshold can be set through the camera control software~\citep{sangiorgi16}, depending on the level of the light of the night sky (LONS).

The ASTRI-Horn telescope has internal~\citep{rodeghiero14,impiombato17} and external~\citep{segreto16} calibration systems, and hardware and software for control~\citep{antolini16} and acquisition~\citep{conforti16}. It also comprises data archiving system~\citep{carosi16} and data reduction and analysis software~\citep{lombardi16}. Furthermore, the station is equipped with various instrumentation devoted to the monitoring of meteorological and environmental conditions~\citep{leto14}. A dedicated control room and a data centre have also been developed as part of the ASTRI Project~\citep{gianotti18}. All data produced during operations are stored in the ASTRI on-site archive, sent to the off-site archive, and analyzed with {\it A-SciSoft}, the official scientific software of the ASTRI Project. The software has been designed to handle both the real and MC data and to provide all necessary algorithms and analysis tools for characterizing the scientific performance of the ASTRI-Horn telescope. {\it A-SciSoft} was designed to handle an array of telescopes, making it suitable for the processing of the data acquired with the ASTRI mini-array~\citep{lombardi18}, proposed as a pathfinder sub-array for the southern site CTA Observatory. \emph{A-SciSoft} is one of the CTA analysis software prototypes to be developed and tested on real data. The aim is to actively contribute to the ongoing efforts for the data handling system of the CTA Observatory.

\section{Observations and analysis}
\label{sec:section3}
The \object{Crab Nebula} (RA (J2000)~=~5h34m31.94s and Dec (J2000)~=~+22$^\circ$00$^\prime$52.2$^{\prime\prime}$) was observed by the ASTRI-Horn telescope over nights between 5 and 11 December 2018 during the verification phase of the system. At the beginning of the campaign, a dedicated trigger threshold scan was performed in order to find the optimal trigger configuration, which resulted in five contiguous pixels above a threshold of 13~pe for a trigger to occur. This configuration provided an average data acquisition rate of the order of 50~Hz. The relatively high trigger threshold adopted for the data taking was due to the rather high level of the LONS at the ASTRI-Horn telescope site~\citep{maccarone13} given the proximity of the city of Catania. The scientific data taking was performed in the so-called ``ON/OFF'' mode~\citep{albert08}, where the data are split in two separate sets. In the so-called ``ON'' data phase, the Crab Nebula was tracked at the center of the FoV, ensuring the maximum camera acceptance. In the so-called ``OFF'' data phase, suitable sky regions without any known gamma-ray source in the FoV were observed (immediately before and after the ON observations) instead across the same range of zenith and azimuth angles as the ON data. The OFF data were then used to properly estimate the level of background in the ON data. All data were taken during dark time at low zenith angles (from $\sim$16$^\circ$ to $\sim$32$^\circ$) for an overall observation time (before data selection) of 25.8~h, evenly divided between ON and OFF observations.

Since the data acquisition was carried out during the telescope verification phase, not all hardware components were already  in their nominal functional state during the scientific observations. The two most relevant sub-systems affected by such suboptimal hardware conditions were the primary mirror and the imaging camera. With regard to the primary mirror, three (out of 18) panels were not in their proper condition. One of them was not adjustable through the active mirror control so the decision was made to cover it with adhesive tape before the data acquisition process. The other two were instead found to be misaligned by $\gtrsim$~1$^{\circ}$ after the data acquisition process by means of a dedicated analysis based on the variance method~\citep{segreto19}. In addition to a systematic decrease of the overall amount of collected Cherenkov light, these panel conditions also potentially affected the shape of the recorded images. This, in turn, could have introduced a possible deterioration on the performance of the shower reconstruction and, thus, on the level of signal detectable with our analysis. However, after a few sanity checks performed on the shape of the recorded images and their main distributions, we concluded that this effect was significantly reduced at the final analysis level. In addition, all panels of the primary mirror had been in operation since 2014 in a harsh environment close to an active volcano, which led to a degradation of their reflectivity. The overall loss of the optical throughput with respect to the nominal condition has been evaluated to be around 30\%. This value has been subsequently confirmed by an independent muon analysis~\citep{mineo19}. For the camera, on the other hand, the high-gain channel (for all camera pixels) was not fully reliable for technical reasons. Therefore, only the low-gain channel~\citep{catalano18} was fiducially considered in the subsequent data reduction. On top of this, 1 (out of 21) PDM at the camera edge was not operative.

All of these hardware issues had some significant impact on the performance of the system. The main effects on the detection of the source were to increase the overall gamma-ray energy threshold (the nominal expected one being around 1 TeV) and to decrease the overall sensitivity of the system. Dedicated MC simulations were performed with the primary aim of generating the gamma/hadron separation look-up-table to be used for the background suppression in the real data~\citep{lombardi17}. All hardware issues mentioned above were considered and implemented in the simulation chain, although some of them could be accounted for with only a limited level of accuracy. While this allowed us to achieve a reasonable match between real and MC-simulated data at the image parameter level (which represents an essential prerequisite for achieving a good background suppression capability), the overall accuracy of our simulations was not high enough to extract spectral parameters for the source. Therefore, we restricted the analysis presented in this work to the production of the detection plot (see Section~\ref{sec:section4}) aimed to obtain a solid evaluation of the signal-to-noise ratio.

For each observing night, the data selection was performed while taking into account several quality checks. First of all, the data affected by technical problems, such as instabilities in the pedestals and signals of the camera PDMs, were rejected, as well as the data showing large fluctuations in the trigger rate and taken during bad atmospheric conditions. The atmospheric conditions (mainly humidity, external temperature, and cloudiness) were monitored by means of dedicated weather station system and an all sky camera. Finally, a stability within 15\% of the data rate at analysis level (i.e., after the cleaning of the shower images and the application of a cut in {\it Size}~$>200$~pe, see later) was considered for the final data selection. The selected data sample resulted in 24.4~h, corresponding to 12.4~h for the ON data and 12.0~h for the OFF data. Table~\ref{tab:1} and Table~\ref{tab:2} summarize the main observational quantities for the ON and OFF selected data samples, respectively.
\begin{table}[ht]
\caption{Date, observation time, zenith angle (Zd) range, and analysis rate of the selected ON data. The analysis rate is calculated after the application of a cut in {\it Size}~$>200$~pe (see text for the definition of {\it size}).}
\label{tab:1}
\begin{center}
\begin {tabular}{|c|c|c|c|}
\hline
Date & Time [h] & Zd~range [deg] & Rate [Hz]\\
\hline
2018.12.05 & 2.0  & 15.7 -- 22.5 & 7.5\\
\hline
2018.12.07 & 2.9  & 15.7 -- 27.5 & 7.7\\
\hline
2018.12.08 & 3.6  & 15.7 -- 31.7 & 7.7\\
\hline
2018.12.09 & 2.5  & 15.7 -- 29.0 & 7.4\\
\hline
2018.12.11 & 1.4  & 15.7 -- 26.4 & 7.0\\
\hline
\hline
Overall    & 12.4 & 15.7 -- 31.7 & 7.5\\
\hline
\end{tabular}
\end{center}
\end{table}
\begin{table}[ht]
\caption{Date, observation time, zenith angle (Zd) range, and analysis rate of the selected OFF data. The analysis rate is calculated after the application of a cut in {\it Size}~$>200$~pe (see text for the definition of {\it size}).}
\label{tab:2}
\begin{center}
\begin {tabular}{|c|c|c|c|}
\hline
Date & Time [h] & Zd~range [deg] & Rate [Hz]\\
\hline
2018.12.05 & 1.6  & 15.8 -- 20.0 & 7.5\\
\hline
2018.12.07 & 3.4  & 15.4 -- 28.1 & 7.6\\
\hline
2018.12.08 & 3.1  & 15.7 -- 27.4 & 7.7\\
\hline
2018.12.09 & 1.4  & 15.7 -- 19.9 & 7.8\\
\hline
2018.12.11 & 2.5  & 15.8 -- 23.0 & 7.2\\
\hline
\hline
Overall    & 12.0 & 15.4 -- 28.1 & 7.6\\
\hline
\end{tabular}
\end{center}
\end{table}

The data reduction and analysis were carried out using {\it A$\mbox{-}$SciSoft}~\citep{lombardi16} in a single-telescope configuration. The raw data, containing the full information available per pixel (integrated signal amplitude in analog-to-digital converter counts) for each triggered shower, were calibrated in order to extract and convert the signal into pe. The conversion coefficients were extracted from specific camera calibration data, taken at the beginning of each observational night. Then, the calibrated data underwent an image cleaning procedure aimed at removing pixels which most likely do not belong to a given Cherenkov shower image. The cleaning method applied to the data was a two-threshold two-pass cleaning. The algorithm first uses a relatively high signal threshold (L1) to search for at least two neighboring pixels (so-called core pixels), which likely belong to the core of the shower. In a second step, pixels adjacent to core pixels (so-called boundary pixels) are included in the cleaned image if their signal is above a lower threshold (L2). For the present analysis, the value of 20~pe for L1 (corresponding to $\gtrsim$~3~times the maximum pedestal RMS present in the data) and 10~pe for L2 (half of the L1 value) were set. The relatively high values of these cleaning thresholds are again the consequence of the rather high level of the LONS at the ASTRI-Horn telescope site. After this step, a parameterization of each cleaned image was performed. The extracted parameters are mainly based on the moments up to the third order of the light distribution on the camera~\citep{hillas85}. Among them, the {\it Size} parameter, defined as the total amount of signal in pe belonging to the cleaned image, was used to compute the average event rate at the analysis level (see Tables~\ref{tab:1} and \ref{tab:2}). In this step, the telescope pointing and the source position (in camera coordinates) were extracted from the data and linked to each parameterized shower image.

After the image parameterization step, several consistency checks on the main Hillas parameters' distributions of the ON and OFF data were performed in order to verify the compatibility of the two data samples after the data selection and before the application of the final analysis cuts. Once the events were cleaned and parameterized, a set of simulated gamma-ray and hadronic background events (treated with the same data processing chain as the real data and reasonably matching the image parameters' distributions of the real data) were used to train a machine learning algorithm based on Breiman's random forest method~\citep{breiman01} for the calculation of a suitable look-up-table for gamma/hadron separation. This step is important for any IACT analysis due to the overwhelming hadronic background. In the present analysis, the image parameters of {\it size}, {\it dist}, {\it width}, {\it length}, and {\it concentration}~\citep{hillas85} were used for training the random forest. Finally, the look-up-table was applied to the ON and OFF data sets in order to get for each event a gamma/hadron discrimination parameter called {\it gammaness}. It ranges from 1 (for showers confidently identified as initiated by gamma rays) to 0 (for those clearly showing the features of a hadronic cosmic-ray initiated shower). At this level, the data were ready for the final step of the analysis, aimed at searching for a gamma-ray excess from the source direction.

\section{Results}
\label{sec:section4}
In order to search for a gamma-ray excess in the data, we compared the $\mid${\it Alpha}$\mid$-distribution of the ON and OFF data (i.e., using the so-called $\mid${\it Alpha}$\mid$-plot) after the application of suitable analysis cuts (mainly in {\it size} and {\it gammaness}) and within a fiducial $\mid${\it Alpha}$\mid$ signal region. {\it Alpha}~\citep{plyasheshnikov85} is the angle between the major axis of the recorded image and the vector connecting its center of gravity with the source position in the camera plane. Gamma-ray shower images from the source tend to point with their major axes toward the source position in the camera, whereas images of background showers do not show any preferred orientation, having isotropic arrival directions. This implies that the $\mid${\it Alpha}$\mid$-distribution for gamma-ray induced showers peaks at low $\mid${\it Alpha}$\mid$ values, whereas the $\mid${\it Alpha}$\mid$-distribution for the background is almost flat across the entire $\mid${\it Alpha}$\mid$ range (from 0$^{\circ}$ to 90$^{\circ}$).

The final analysis cuts were selected by means of dedicated MC gamma-ray and real background data. A scan procedure aimed at finding the best sensitivity to gamma-ray excess (assuming a Crab-Nebula-like spectrum,~\citealt{albert08}) as a function of different combination of cuts in {\it size}, {\it gammaness}, and $\mid${\it Alpha}$\mid$ parameters was employed. As a result, the optimal cuts {\it Size}~$>300$~pe, {\it gammaness}~$>0.88$, and $\mid${\it Alpha}$\mid$~$\leq10^{\circ}$ were found and applied to real data. This set of cuts corresponds to an energy threshold (defined as the peak of the MC gamma-ray simulated energy distribution for a Crab-Nebula-like spectrum after all analysis cuts) of $\sim$3~TeV.

\begin{figure}
\centering
\includegraphics[width=0.5\textwidth]{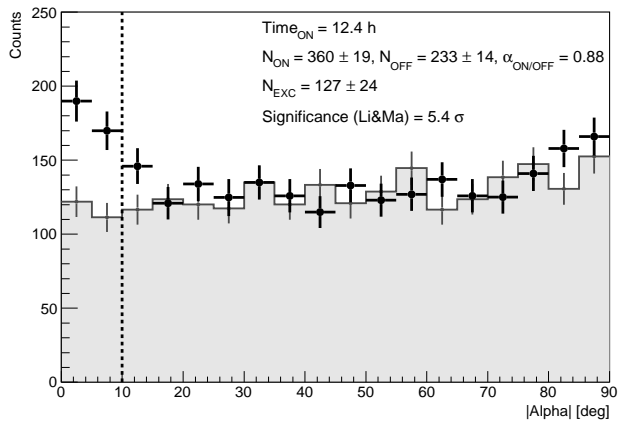}
\caption{$\mid${\it Alpha}$\mid$-distributions of the Crab Nebula (ON, black) and the background (OFF, grey) data from ASTRI-Horn observations taken between 5 and 11 December 2018 above an energy threshold of $\sim$3~TeV. The region between zero and the vertical dashed line (at 10$^{\circ}$) represents the fiducial signal region.}
\label{fig:figure2}
\end{figure}

The $\mid${\it Alpha}$\mid$-plot achieved from the ON and OFF selected data, following the application of the optimized analysis cuts listed above, is shown in Fig.~\ref{fig:figure2}. The distribution of the OFF data was scaled by a factor $\alpha_{\mbox{\scriptsize ON/OFF \normalsize}}$ found by normalizing the $\mid${\it Alpha}$\mid$-distributions of both samples between 20$^{\circ}$ and 80$^{\circ}$. We found an excess of 127$\pm$24 events, corresponding to a significance of 5.4 standard deviations ($\sigma$), calculated according to Eq.~17 of~\cite{lima83}. It should be noted that the energy threshold of $\sim$3~TeV quoted above is substantially in line with the actual number of detected gamma-ray events. In fact, assuming a typical radius of $\sim$125~m for the Cherenkov light pool at $\sim$2000~m a.s.l.~\citep{denaurois15} (corresponding to a gamma-ray effective area of $\sim$5$\times$10$^4$~m$^2$) and taking also into account the fact that the large FoV of the ASTRI camera (7.6$^{\circ}$) allows for the detection of gamma-rays with even larger impact parameters, we can estimate that an energy threshold of $\gtrsim$~2.5~TeV would provide a number of excesses from the Crab Nebula~\citep{aharonian06,aleksic16} comparable with the actual detected ones, in the given exposure time (12.4~h).

\section{Conclusions}
\label{sec:section5}
In this work, we report on the first detection at the VHE of the Crab Nebula by a Cherenkov telescope in dual-mirror Schwarzschild-Couder configuration: the ASTRI-Horn telescope. The telescope, installed on Mt. Etna in Italy, is part of the ASTRI Project, led by the INAF in the context of prototyping the small-size class of telescopes of the Cherenkov Telescope Array Observatory.

The ASTRI-Horn telescope is one of the first three dual-mirror Cherenkov telescope prototypes adopting technological innovations, such as the dual-mirror aplanatic Schwarzschild-Couder optical configuration and a wide field compact SiPM camera with a very fast response, high photon detection efficiency, and excellent single photo-electron resolution. All the adopted innovative solutions have been tested during an observational campaign on the Crab Nebula carried out in December 2018 as part of the verification phase of the telescope. The acquired data have been reduced and analyzed using {\it A-SciSoft}, the official ASTRI scientific software package developed as part of the ASTRI Project. During observations, the hardware status of the system was not yet in its nominal condition, which prevented the system from operating at the nominal performance level. Nevertheless, data analysis has led to the detection of the source at a statistical significance of 5.4~$\sigma$ above an energy threshold of $\sim$3~TeV in 12.4~h of on-axis observations.

Although no spectral parameters of the source were derived from the present analysis, this result nonetheless represents an important step towards the validation of the dual-mirror optical design for ground-based gamma-ray astronomy applications. The dual-mirror optical design for Cherenkov telescopes is a particularly attractive solution because it makes a good angular resolution across the entire field of view possible. Indeed, the dual-mirror solution enables a better correction of aberrations at large field angles and, hence, the construction of telescopes with a smaller focal ratio, allowing the use of compact cameras able to cover with mm-sized pixels a large field of view.

The scientific validation phase of the ASTRI-Horn telescope is foreseen to start in spring 2020, hosting new, extensive campaigns on a few bright gamma-ray sources, including the Crab Nebula. In view of this phase, some hardware improvements (mainly on the optical system and camera) are already scheduled in order for it to reach the nominal configuration and, consequently, the best performance. This will eventually allow us to fully characterize the hardware innovations of the system and to set the path towards their systematic implementation for the next generation of Cherenkov telescopes.

\begin{acknowledgements}
This work was conducted in the context of the CTA ASTRI Project. This work is supported by the Italian Ministry of Education, University, and Research (MIUR) with funds specifically assigned to the Italian National Institute of Astrophysics (INAF) for the Cherenkov Telescope Array (CTA), and by the Italian Ministry of Economic Development (MISE) within the ``Astronomia Industriale'' program. We acknowledge support from  the Brazilian Funding Agency FAPESP (Grant 2013/10559-5) and from the South African Department of Science and Technology through Funding Agreement 0227/2014 for the South African Gamma-Ray Astronomy Programme. We gratefully acknowledge financial support from the agencies and organizations listed here: {\tt http://www.cta-observatory.org/consortium\_acknowledgments}.
This work has been supported by H2020-ASTERICS, a project funded by the European Commission Framework Programme Horizon 2020 Research and Innovation action under grant agreement n. 653477.\\
This work has gone through internal review by the CTA Consortium.
\end{acknowledgements}

\end{document}